\documentclass[a4paper,11pt,reqno]{amsart}
\usepackage{hyperref}
\usepackage{graphicx}
\usepackage{multicol}
\usepackage{color}
\usepackage{caption}
\usepackage{setspace}
\usepackage{amssymb,amsmath}  
\numberwithin{equation}{section}   
\usepackage{amsfonts,amsrefs}
\usepackage{amsthm} 
\usepackage{bm}    
\usepackage{bbm}      
\usepackage{tabularx}
\usepackage{enumerate}
\usepackage[linesnumbered,ruled,vlined]{algorithm2e}
\usepackage{algorithmic}
\usepackage{lscape}
\makeatletter
\def\verbatim@font{\linespread{1}\normalfont\ttfamily}
\makeatother
\usepackage{pgfpages}
\usepackage{changepage}
\usepackage{fancyhdr,a4wide}
\usepackage{stmaryrd}

\usepackage[utf8]{inputenc}
\usepackage[T1]{fontenc}
\usepackage{amsmath}
\usepackage{tensor}
\usepackage{amssymb}
\usepackage{graphicx}
\usepackage{subfigure}
\usepackage{mathrsfs}
\usepackage{microtype}
\usepackage{rotating}
\usepackage{xspace}
\usepackage{mathtools}
\usepackage{tikz,pgfplots}
\usepackage{comment}
\usepackage{MnSymbol}
\pgfplotsset{compat=1.9}
\usetikzlibrary{fit}
\tikzset{every label/.style={font=\footnotesize,inner sep=1.5pt}}

\allowdisplaybreaks
\def\jf#1{\relax}

\newcommand*{\del}{\partial}

\theoremstyle{plain}
\newtheorem{thm}{Theorem}[section]

\theoremstyle{definition}

\theoremstyle{remark}
\newtheorem{rem}[thm]{Remark}

\usepackage{placeins}
\usepackage{multirow}

\newcommand*{\R}[1]{\mbox{\sffamily\ifcase#1\relax\or I\or II\or III\or IV\fi}\xspace}

\title[Instability of Slowly Expanding FLRW Spacetimes]{Instability of Slowly Expanding FLRW Spacetimes}
\author[E.~Marshall]{Elliot Marshall}
\address{School of Mathematics\\
9 Rainforest Walk\\
Monash University, VIC 3800\\ Australia}
\email{elliot.marshall@monash.edu}

\begin{document}

\begin{abstract} 
We numerically study, under a Gowdy symmetry assumption, nonlinear perturbations of the decelerated FLRW fluid solutions to the Einstein-Euler system toward the future for linear equations of state $p=K\rho$ with $0\leq K\leq 1$. This article builds on the work of Fajman et al. \cite{Fajman_et_al:2024} in which perturbations of the homogeneous fluid solution on a fixed, decelerating FLRW background were studied. Our numerical results show that for all values of $K$, perturbations of the FLRW solution develop shocks in finite time. This behaviour contrasts known results for spacetimes with accelerated expansion in which shock formation is suppressed. 
\end{abstract}

\maketitle

\section{Introduction}
It is well-known, due to a celebrated result of Christodoulou \cite{Christodoulou:2007}, that relativistic fluids on Minkowski space generically form shocks in finite time. In cosmological models, however, it has been shown that spacetime expansion can suppress shock formation in fluids, a process called \textit{fluid stabilisation}. This was originally proven in the Newtonian case by Brauer, Rendall, and Reula \cite{RendallBrauerReula:1994}. The first rigorous results in the relativistic setting were established by Rodnianski and Speck \cites{RodnianskiSpeck:2013,Speck:2012} who proved the future stability of nonlinear perturbations of Friedmann-Lemaître-Robertson-Walker (FLRW) solutions to the Einstein-Euler equations with a positive cosmological constant, $\mathbb{T}^{3}$ spatial topology, and linear equations of state $p=K\rho$ for the parameter range $0 < K < 1/3$. Subsequent work has established the stability of models with $K=0$ \cite{HadzicSpeck:2015} and $K=\frac{1}{3}$ (with $\mathbb{S}^{3}$ spatial topology) \cite{LubbeKroon:2013}, hyperbolic spatial slices (for $0<K<\frac{1}{3}$) \cites{Mondal:2021,Mondal:2024}, and alternative equations of state \cites{LiuWei:2021,LeFlochWei:2021}.  \newline \par

As the prototypical example of an expanding universe, FLRW spacetimes are the simplest models in which to study the role of spacetime expansion on fluid stabilisation. To this end, let us consider a generic FLRW-type metric
\begin{align}
    \label{eqn:FLRW_Example}
    g = -d\tau^{2} + a(\tau)^{2}h_{ij}dx^{i}dx^{j},
\end{align}
where $h_{ij}$ is a Riemannian metric on a compact manifold without boundary. The scale factor $a(\tau)$ is an increasing function of the time $\tau$ and $\tau=\infty$ is future timelike infinity. We can classify the rate of expansion for the metric \eqref{eqn:FLRW_Example} as follows
\begin{itemize}
    \item Accelerated Expansion - $a^{\prime\prime}(\tau) >0$.
    \item Linear Expansion - $a^{\prime\prime}(\tau) = 0$.
    \item Decelerated Expansion = $a^{\prime\prime}(\tau) <0$.
\end{itemize}
On a heuristic level, the expectation is that a faster rate of expansion is more effective at suppressing shocks. Numerous rigorous results, including the aforementioned work of Rodnianski and Speck, have been established on the stability of fluid-filled spacetimes with accelerated expansion \cites{HadzicSpeck:2015,LubbeKroon:2013,Oliynyk:CMP_2016,Wei:2018,LeFlochWei:2021,Fournodavlos_et_al:2024,MarshallOliynyk:2022,Oliynyk:2021}, however significantly less is known in the case of linear or decelerated expansion. The purpose of this article is to numerically investigate shock formation for small perturbations of decelerated FLRW solutions to the Einstein-Euler equations. In particular, we provide convincing evidence that perturbed solutions \textit{always} form shocks in finite time. A brief overview of previous stability results in the linear and decelerated regimes is given in the following section. \newline \par

\subsection{Previous Results for Linear and Decelerated Expansion}
We say a solution to the relativistic Euler or Einstein-Euler equations is \textit{stable} if small perturbations exist globally to the future and remain close to the background solution (in some appropriate sense) and \textit{unstable} otherwise. In the context of this article, we focus on the case where instabilities are due to the formation of shocks in the fluid matter. For concreteness, let us specialise to the case of the relativistic Euler equations with linear equation of state\footnote{Here $K=c_{s}^{2}$ is the square of the sound speed.} $p=K\rho$ ($0\leq K \leq 1$) on a fixed background with a power-law scale factor $a(\tau)=\tau^{\sigma}$, $\sigma>0$. For this particular scale factor, $\sigma>1$ corresponds to accelerated expansion while $\sigma<1$ is decelerated expansion.  \newline \par

\begin{figure}
    \centering
    \includegraphics[width=0.5\textwidth]{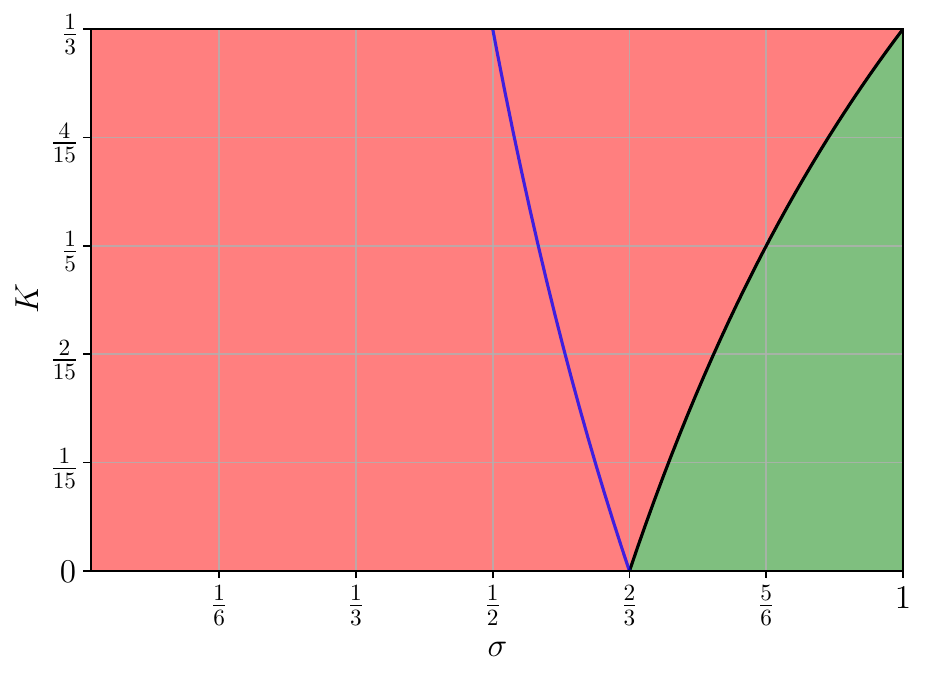}
    \caption{Plot of the $(\sigma,K)$ parameter space for fluids on a power-law background established in \cite{Fajman_et_al:2024}. The green and red shaded regions denote the stable and unstable regimes, respectively. The critical line, $K_{\text{crit}}=1-\frac{2}{3\sigma}$ is coloured black while the blue line denotes the FLRW solutions to the Einstein-Euler system with zero cosmological constant.}
    \label{fig:Parameter_SpacePlot}
\end{figure}

In the case of dust ($K=0$), Speck \cite{Speck:2013} established stability for $\sigma>\frac{1}{2}$. For the case $\sigma=1$, stability for irrotational fluids with $K\in(0,1/3)$ was proven by Fajman, Oliynyk, and Wyatt \cites{FOW:2021}. The irrotational restriction was subsequently removed in a recent article by Fajman, Ofner, Oliynyk, and Wyatt \cite{FOOW:2023}. Shock formation for $\sigma=1$ at the endpoint $K=1/3$ was also established by Speck in \cite{Speck:2013}. Recently, Fajman et al. \cites{Fajman_et_al:2024, Fajman_et_al:2025} provided analytic and numerical evidence of a transition between stable and unstable fluids on fixed decelerated spacetimes. The main results from \cites{Fajman_et_al:2024,Fajman_et_al:2025} can be summarised as follows: 
\begin{itemize}
\item Dust fluids ($K=0$) are unstable for $\sigma\leq1/2$ and stable otherwise.
\item Barotropic fluids ($K>0$) are unstable when $\sigma<\frac{2}{3}$.  
\item For $\sigma \in (2/3,1)$, the $(K,\sigma)$ parameter space is split along the critical line $K_{\text{crit}}=1-\frac{2}{3\sigma}.$ Fluids with $0<K<K_{\text{crit}}$ are stable, while fluids with $K>K_{\text{crit}}$ form shocks (cf. Figure \ref{fig:Parameter_SpacePlot}). 
\end{itemize}
Importantly, the $\mathbb{T}^{3}$ FLRW metrics for $K>0$ lie entirely within the \textit{unstable} regime predicted in \cite{Fajman_et_al:2024}. On the other hand, the dust FLRW spacetime ($K=0$) lies in the \textit{stable} regime. Finally, we note for the \textit{coupled} Einstein-Euler system\footnote{Note, however, a recent paper by Taylor \cite{Taylor:2023} which proves future stability of decelerated FLRW solutions to the Einstein-Massless Vlasov system with \textit{non-compact} spatial slices.} there are only two known rigorous stability results \cites{Fajman_et_al:2021b, FOOW:2023}, both of which consider perturbations of background spacetimes with linear expansion and negative spatial curvature. 

\subsection{Outline of Paper}
The main aim of this article is to numerically investigate the behaviour of non-linear perturbations of the \textit{decelerated} FLRW solution to the \textit{coupled} Einstein-Euler system for the full sound speed range $0 \leq K \leq 1$. This extends the work of Fajman et al. \cite{Fajman_et_al:2024} by including coupling to the gravitational field. In particular, we are interested in studying the formation of shocks from small data perturbations. To this end, we restrict to the Gowdy symmetric Einstein-Euler equations with $\mathbb{T}^{3}$ spatial topology. Gowdy symmetry has the benefit of reducing the Einstein equations to a $(1+1)$-dimensional problem with periodic boundary conditions. This reduction has previously been used to great effect in numerous analytic and numerical studies \cites{ames2017,amorim2009,Berger:1993,BeyerHennig:2012,beyer2010b,beyer2017,beyer2020b,chrusciel1990,isenberg1990,kichenassamy1998,LeFlochRendall:2011,rendall2000,ringstrom2009a,BMO:2023,BMO:2024}. \newline \par

A key limitation of the numerical study in \cite{Fajman_et_al:2024} was that their scheme could not stably evolve shocks in the fluid. Indeed, while their results provide convincing evidence that shocks will imminently form they do not actually evolve the fluid to the point that a shock forms. In contrast, we evolve the fluid using the high-resolution shock capturing Kurganov-Tadmor scheme \cite{KurganovTadmor:2000}. Importantly, this allows us to evolve \textit{past} the point of shock formation and stably model shock waves in the numerical domain.\newline \par

Our numerical simulations yield the following observations:
\begin{enumerate}[(a)]
    \item For sound speeds $K \in (0,1]$ and initial data suitably close to FLRW initial data, highly oscillatory shocks form in the fluid variables. The gravitational variables also display highly oscillatory behaviour which does not appear to settle down at late times. 
    \item For $K>0$, the shock formation time $t_{*}$ decreases exponentially as the size of the perturbation and $K$ are increased. In particular we find, for fixed $K$, $t_{*}$ scales like  $A\exp(\frac{B}{\|I.D.\|_{2}^{2}} + \frac{C}{\|I.D.\|_{2}})$ where $\|I.D.\|_{2}$ is the norm of the initial data (cf. Section \ref{sec:shock_time_tests}).
    \item For $K=0$ and initial data suitably close to FLRW initial data, stationary shocks appear to form in the fluid velocity. This contrasts the behaviour of dust fluids on fixed backgrounds with $\sigma>\frac{1}{2}$, which are stable.
\end{enumerate} 

\begin{rem}
    In the work of Brauer, Rendall, and Reula \cite{RendallBrauerReula:1994}, it was shown that characteristic curves cross in finite time for perturbations of the decelerated dust FLRW model ($a(\tau)=\tau^{\frac{2}{3}}$) in Newtonian gravity. That is to say, perturbed solutions break down in finite time. On the other hand, as discussed above, the results of Speck \cite{Speck:2013} show that (relativistic) dust fluids are stable on fixed backgrounds with  $a(\tau)=\tau^{\sigma}$, $\sigma>\frac{1}{2}$. Together these results suggest that fluids are less stable when coupled to the gravitational field. Indeed, this interpretation is consistent with our results for perturbed dust FLRW spacetimes (see point (c) above).
\end{rem}

The article is structured as follows: the derivation of a flux-conservative form of the Gowdy-symmetric Einstein-Euler equations suitable for numerical implementation is carried out in Section \ref{sec:Einstein_Derivation}. In Section \ref{sec:FLRW_Derivation} we derive the FLRW solution in our variables. Finally, in Section \ref{sec:NumericalResults}, we discuss our numerical setup and results.

\section{Einstein-Euler Equations}
\label{sec:Einstein_Derivation}
\subsection{Einstein-Euler field equations with Gowdy symmetry}
\label{sec:EinsteinEulerderivationGowdy}
The Einstein-Euler equations\footnote{Our indexing conventions are as follows: lower case Latin letters, e.g. $i,j,k$,
will label spacetime coordinate indices that run from $0$ to $3$ while upper case Latin letters, e.g. $I,J,K$, will label spatial coordinate indices that run from
$1$ to $3$.} for a perfect fluid are given by\footnote{Here, we use units where $c=1$ and $G=\frac{1}{8\pi}$.} 
\begin{align}
\label{eqn:Einstein1}
    G_{ij} &=T_{ij}, \\
    \label{eqn:Tij_divergence}
\nabla_{i}T^{ij} &= 0, 
\end{align}
where
\begin{align*}
    T_{ij} &= (\rho+p)u_{i}u_{j}+pg_{ij} 
\end{align*}
is the perfect fluid stress-energy tensor. Here, $u_{i}$ is the fluid four-velocity normalised by $u_{i}u^{i}=-1$, and we assume that the fluid's proper energy density, $\rho$, and pressure, $p$, are related via the linear equation of state 
\begin{align*}
    p = K\rho,
\end{align*}
where the constant parameter $K\geq 0$ is the square of the sound speed. In the following, we assume that $0\leq K \leq 1$ so that the speed of sound is less than or equal to the speed of light.

As discussed in the introduction, we restrict our attention to solutions of the Einstein-Euler equations with a Gowdy symmetry \cites{chrusciel1990,gowdy1974} by
considering Gowdy metrics in areal coordinates on $\mathbb{R}_{>0} \times \mathbb{T}^{3}$ of the form
\begin{align}
\label{eqn:gowdymetric}
    g = e^{2(\eta-U)}(-e^{2\bar{\alpha}}d\bar{t}^{2} + dx^{2}) + e^{2U}(dy^{2}+Adz^{2}) + e^{-2U}\bar{t}^{2}dz^{2}.
\end{align}
Here, the functions $\eta$, $U$, $\bar{\alpha}$, and $A$ depend only on $(\bar{t},x)\in \mathbb{R}_{>0}\times \mathbb{R}$ and are $2\pi$-periodic in $x$. These coordinates are called `areal' because the time coordinate $\bar{t}$ is proportional to the area of the symmetry orbits labelled by the coordinates $y$ and $z$. In particular we have, 
\begin{align}
    \det\begin{pmatrix} e^{2U} & Ae^{2U} \\ Ae^{2U} & e^{2U}A^{2} + e^{-2U}\bar{t}^{2} \end{pmatrix} = \bar{t}^{2}.
\end{align}
It is natural to work with areal coordinates as they provide a \textit{global foliation}\footnote{That is, a foliation which covers the maximal globally hyperbolic development for a given family of spacetimes.} for Gowdy-symmetric Einstein-Euler spacetimes \cite{LeFlochRendall:2011} (see also \cites{BCIM:1997,Moncrief:1981}). Inspired by the work of LeFloch and collaborators on low-regularity solutions to the Einstein-Euler equations in Gowdy symmetry \cites{LeFlochRendall:2011,GrubicLeFloch:2013,GrubicLeFloch:2015}, we introduce a new time coordinate $t$ (note that we still have an areal foliation) and metric functions $\alpha$ and $\nu$ via
\begin{align}
    \bar{t} = e^{t}, \;\; \bar{\alpha} = \alpha -t, \;\; \nu = \eta+\alpha.
\end{align}
In terms of these variables, the Gowdy metric is expressed as
\begin{align}
\label{eqn:gowdymetricA}
    g = e^{2(\nu-U)}(-dt^{2} + e^{-2\alpha}dx^{2}) + e^{2U}(dy+Adz)^{2} + e^{-2U+2t}dz^{2}.
\end{align}
As we are only interested in solutions in the expanding
direction, i.e. towards the future, we consider the time interval of  $t \in [0, \infty)$. We are now in a position to express the Gowdy-symmetric Einstein-Euler equations in a first-order form suitable for numerical implementation.

\subsection{A First Order Formulation of the Einstein Equations}
In Gowdy symmetry, the fluid four-velocity only has two non-zero components\footnote{This follows from choosing coordinates where the two Killing vectors are given by $\del_{y}$ and $\del_{z}$, see \cite{LeFlochRendall:2011}.} and can be expressed as
\begin{equation} \label{u-Gowdy}
u=u_0 dt  + u_1 dx,
\end{equation}
where the functions $u_0$ and $u_1$ depend on $(t,x)\in \mathbb{R}\times \mathbb{R}$ and are $2\pi$-periodic in $x$. Following \cite{LeFlochRendall:2011}, we define the \textit{scalar velocity} by
\begin{align}
    v = \frac{u^{1}}{e^{\alpha}u^{0}}, \;\; |v| \in [0,1).
\end{align}
Using the normalisation of the four-velocity we obtain the identity
\begin{align}
    (u^{0})^{2} - e^{-2\alpha}(u^{1})^{2} = e^{-2(\nu-U)}.
\end{align}
A short computation then yields
\begin{align}
    e^{2(\nu-U)}(u^{0})^{2} = \frac{1}{1-v^{2}}, \\
    e^{2(\nu-U-\alpha)}(u^{1})^{2} = \frac{v^{2}}{1-v^{2}}.
\end{align}
Next, we define the modified density, $\mu$, by 
\begin{align}
\label{eqn:mu_defn}
   \mu := e^{2(\nu-U)}\rho.
\end{align}
In terms of $v$ and $\mu$, the non-vanishing components of the stress-energy tensor are given by
\begin{align*}
T_{00} &= (1+K)\mu \Gamma^{2}- K\mu, \\
T_{01} &= -e^{-\alpha}(1+K)\mu\Gamma^{2}v, \\
T_{11} &= e^{-2\alpha}\mu\big(K+(1+K)\Gamma^{2}v^{2}\big), \\ 
T_{22} &= e^{4U-2\nu}K\mu, \\
T_{23} &= e^{4U-2\nu}AK\mu, \\ 
T_{33} &= e^{-2\nu}(e^{2t}+e^{4U}A^{2})K\mu,
\end{align*}
where 
\begin{align}
    \Gamma = \frac{1}{\sqrt{1-v^{2}}}.
\end{align}
Using these  expressions and the Gowdy metric \eqref{eqn:gowdymetricA}, 
a straightforward calculation shows that the Einstein equation \eqref{eqn:Einstein1} in Gowdy symmetry consists of the following three wave equations
\begin{align}
\label{eqn:Awave}
\del_{tt}A &= e^{2\alpha}\Big(A_{x}\big(4U_{x}+\alpha_{x}\big) + A_{xx}\Big) + A_{t}(1-4U_{t}+\alpha_{t}), \\
\label{eqn:Uwave}
\del_{tt}U &= \frac{1}{2}\Big(2e^{2\alpha}\big(U_{x}\alpha_{x}+U_{xx}\big) + e^{-2t+4U}\big(-e^{2\alpha}A_{x}^{2}+A_{t}^{2}\big) + \big(-1+2U_{t}\big)\big(-1+\alpha_{t}\big)\Big) , \\
\label{eqn:Nuwave}
\del_{tt}\nu &= -K\mu + e^{2\alpha}\big(U_{x}^{2} + \alpha_{x}\nu_{x} + \nu_{xx}\big) + \frac{1}{4}e^{-2t+4U}\big(-e^{2\alpha}A_{x}^{2} + A_{t}^{2}\big) - U_{t}^{2} \nonumber \\
&+ \alpha_{t}\big(-\alpha_{t}+\nu_{t}\big) + \alpha_{tt}, 
\end{align}
and three first-order equations
\begin{align}
\label{eqn:alpha_evo}
\del_{t}\alpha &= 1 + (K-1)\mu, \\
\label{eqn:hamiltonian_constraint}
\del_{t}\nu &= 1 + \frac{(K+v^{2})\mu}{1-v^{2}} + e^{2\alpha}U_{x}^{2} + \frac{1}{4}e^{-2t+4U}\big(e^{2\alpha}A_{x}^{2} + A_{t}^{2}\big) + U_{t}^{2}, \\
\label{eqn:momentum_constraint}
\del_{x}\nu &= \frac{e^{-\alpha}(1+K)v\mu}{-1+v^{2}} + \frac{1}{2}e^{-2t+4U}A_{x}A_{t} + 2U_{x}U_{t},
\end{align}
where \eqref{eqn:hamiltonian_constraint} and \eqref{eqn:momentum_constraint} are the Hamiltonian and momentum constraints respectively. We note that $\nu$ does not appear in the evolution equations for the variables $A$, $U$, and $\alpha$ or in the evolution equations for the fluid. Thus, we can instead evolve the reduced geometric system consisting only of \eqref{eqn:Awave}-\eqref{eqn:Uwave} and \eqref{eqn:alpha_evo}. As shown in \cite{GrubicLeFloch:2013}, it is sufficient to solve for these variables alone without initially imposing the constraint \eqref{eqn:hamiltonian_constraint}. Recall, however, that the $\mathbb{T}^{3}$ spatial topology means that the metric variables must be $2\pi$-periodic. This implies that the integral of \eqref{eqn:momentum_constraint} vanishes,
    \begin{align}
    \label{eqn:integral_condition_ID}
        \int_{0}^{2\pi} \Big(\frac{e^{-\alpha}(1+K)v\mu}{-1+v^{2}} + \frac{1}{2}e^{-2t+4U}A_{x}A_{t} + 2U_{x}U_{t} dx \Big) = 0,
    \end{align}
which must be imposed on our initial data. Given a solution $(A, U,\alpha,v,\mu)$ to \eqref{eqn:Tij_divergence}, \eqref{eqn:Awave}-\eqref{eqn:Uwave}, and \eqref{eqn:alpha_evo}, $\nu$ can then be recovered by solving the constraint equations\footnote{See Section 3.2 of \cite{GrubicLeFloch:2013} for details.}. Thus, in our numerical scheme we will ignore the equations for $\nu$. \newline \par

Now, by introducing the first order variables
\begin{align}
\label{eqn:firstordervariablesA_U}
A_{0} &= \del_{t}A, \quad A_{1} = e^{\alpha}\del_{x}A, \quad U_{0} = \del_{t}U, \quad U_{1} = e^{\alpha}\del_{x}U,
\end{align}
we can express the wave equations \eqref{eqn:Awave}-\eqref{eqn:Uwave} for $A$ and $U$ in flux conservative form
\begin{align}
\label{eqn:A0_FC_evo}
    \del_{t}\Big(e^{-\alpha}A_{0}\Big) - \del_{x}\Big(A_{1}\Big) &= e^{-\alpha}\big(A_{0} + 4A_{1}U_{1} - 4A_{0}U_{0}\big), \\
\label{eqn:A1_FC_evo}
    \del_{t}\Big(e^{-\alpha}A_{1}\Big) - \del_{x}\Big(A_{0}\Big) &= 0, \\
\label{eqn:U0_FC_evo}
    \del_{t}\Big(e^{-\alpha}(U_{0}-\frac{1}{2})\Big) - \del_{x}\Big(U_{1}\Big) &= e^{-\alpha}\big(\frac{1}{2}e^{-2t+4U}(A_{0}^{2}-A_{1}^{2}) +\frac{1}{2} - U_{0}\big) , \\
\label{eqn:U1_FC_evo}
    \del_{t}\Big(e^{-\alpha}U_{1}\Big) - \del_{x}\Big(U_{0}\Big) &= 0.
\end{align}

\begin{rem}
It is straightforward to see that the flux conservative equations \eqref{eqn:A0_FC_evo}-\eqref{eqn:U1_FC_evo} can also be recast as the following symmetric hyperbolic system,
\begin{align}
\label{eqn:Asymhyp}
\del_{t}\begin{pmatrix} A_{0} \\ A_{1} \end{pmatrix} + \begin{pmatrix} 0 & -e^{\alpha} \\ -e^{\alpha} & 0 \end{pmatrix} \del_{x}\begin{pmatrix} A_{0} \\ A_{1} \end{pmatrix}  -\alpha_{0}\begin{pmatrix} A_{0} \\ A_{1} \end{pmatrix} =& \begin{pmatrix} A_{0} + 4A_{1}U_{1} - 4A_{0}U_{0}   \\ 0 \end{pmatrix}, \\
\label{eqn:Usymhyp}
\del_{t}\begin{pmatrix} U_{0} \\ U_{1} \end{pmatrix} + \begin{pmatrix} 0 & -e^{\alpha} \\ -e^{\alpha} & 0 \end{pmatrix} \del_{x}\begin{pmatrix} U_{0} \\ U_{1} \end{pmatrix}  -\alpha_{0}\begin{pmatrix} U_{0} \\ U_{1} \end{pmatrix} =& \begin{pmatrix} \frac{1}{2}e^{-2t+4U}(A_{0}^{2}-A_{1}^{2}) +\frac{1}{2} - U_{0} -\frac{1}{2}\alpha_{0}\\ 0 \end{pmatrix}.
\end{align}
\end{rem}

\subsection{Conservative Form of the Relativistic Euler Equations}
\label{sec:conservation_euler}
In order to stably evolve discontinuities in the fluid, we must express the Euler equations as a system of balance laws. First, expanding the Euler equations \eqref{eqn:Tij_divergence} yields
\begin{align}
    \del_{t}(\sqrt{-g}T^{0b}) + \del_{x}(\sqrt{-g}T^{1b}) = -\sqrt{-g}\tensor{\Gamma}{_a^b_c}T^{ac}.
\end{align}
where 
\begin{align}
    \sqrt{-g} = e^{t-2U-\alpha+2\nu}.
\end{align}
In terms of $v$ and $\mu$, the relevant components of the stress-energy tensor can be expressed as follows
\begin{align}
\sqrt{-g}T^{00} &= e^{t-\alpha-2\nu+2U}\big[(K+1)\mu\Gamma^{2} -K\mu\big], \\
\sqrt{-g}T^{01} &= e^{t-2\nu+2U}\big[(K+1)\mu\Gamma^{2}v\big], \\
\sqrt{-g}T^{11} &= e^{t+\alpha-2\nu+2U}\big[(K+1)\mu\Gamma^{2}v^{2} +K\mu\big].
\end{align}
After substituting in the constraints \eqref{eqn:hamiltonian_constraint}-\eqref{eqn:momentum_constraint} and the evolution equation for $\alpha$ \eqref{eqn:alpha_evo}, we obtain the following form of the Euler equations
\begin{align}
\label{eqn:Euler_FC_1}
    &\del_{t}\Big(e^{-\alpha}[(K+1)\mu\Gamma^{2} -K\mu]\Big) + \del_{x}\Big([(K+1)\mu\Gamma^{2}v]\Big)  \nonumber \\
    &=\frac{1}{4} (K-1) e^{-\alpha -2t} \mu  \Big(-e^{4 U} (A_{1}^2 +A_{0}^2)-4 e^{2 t} \big( U_{1}^2+(U_{0}-1) U_{0}
   \big)\Big)\nonumber \\ 
   &-e^{-\alpha }K\mu , \\
\label{eqn:Euler_FC_2}
    &\del_{t}\Big(e^{-\alpha}[(K+1)\mu\Gamma^{2}v]\Big) + \del_{x}\Big([(K+1)\mu\Gamma^{2}v^{2} +K\mu]\Big) \nonumber \\
    &= e^{-\alpha}\mu\Big(\frac{1}{2} (K-1) A_{1} A_{0} e^{4 U-2t}+ (K-1)U_{1}
   \left(2 U_{0}-1\right)\Big).
\end{align} 
Observe that the source terms in \eqref{eqn:Euler_FC_1}-\eqref{eqn:Euler_FC_2} do not contain any derivatives of $\alpha$ and $\nu$. 
The fluid equations can be expressed in matrix form by
\begin{align}
\label{eqn:Balance_LawSystem}
    \del_{t}V + \del_{x}F = G
\end{align}
with
\begin{align}
    V &:= \begin{pmatrix} e^{-\alpha}\tau \\ e^{-\alpha}S  \end{pmatrix}, \\
    F &:= \begin{pmatrix} S \\   H \end{pmatrix}, \\
    G &:= \begin{pmatrix}\frac{1}{4} (K-1) e^{-\alpha -2t} \mu  \Big(-e^{4 U} (A_{1}^2 +A_{0}^2)-4 e^{2 t} \big( U_{1}^2+(U_{0}-1) U_{0}
   \big)\Big)-e^{-\alpha }K\mu \\ e^{-\alpha}\mu\Big(\frac{1}{2} (K-1) A_{1} A_{0} e^{4 U-2t}+ (K-1)U_{1}
   \left(2 U_{0}-1\right)\Big) \end{pmatrix}.
\end{align}
where
\begin{align}
    \tau &:= (K+1)\mu\Gamma^{2} -K\mu, \\ 
    S &:= (K+1)\mu\Gamma^{2}v, \\
    H &:= (K+1)\mu\Gamma^{2}v^{2} +K\mu.
\end{align}
The primitive variables $\mu$ and $v$ can be recovered using the following algebraic relations,
\begin{align}
\label{eqn:Gamma_recovery}
    \Gamma^{2} &= \frac{1-2K(1+K)Q + \sqrt{1-4KQ}}{2(1-(1+K)^{2}Q)} , \\
\label{eqn:z_recovery}
    \mu &= \frac{\tau}{(K+1)\Gamma^{2}-K}, \\
\label{eqn:w_recovery}
    v &= \frac{S}{(K+1)\Gamma^{2}\mu},
\end{align}
where 
\begin{align}
    Q := \frac{S^{2}}{(1+K)^{2}\tau^{2}}.
\end{align}

\subsection{Characteristic Structure of the Relativistic Euler Equations}
In order to calculate the characteristic speeds of the Euler equations, we will re-express the balance laws \eqref{eqn:Euler_FC_1}-\eqref{eqn:Euler_FC_2} as a quasi-linear system in terms of the primitive variables $W = (\mu,v)$. This system is given by\footnote{Note, only the principal part is relevant for this analysis.}
\begin{align}
\label{eqn:Primitive_QuasiLinear}
    A^{0}\del_{t}W + A^{1}\del_{x}W = 0,
\end{align}
where
\begin{align*}
    A^{0} &= \frac{\del V}{\del W} = \begin{pmatrix}
- \frac{1 + K v^2}{e^{\alpha} (-1 + v^2)} & \frac{2 (1 + K) v \mu}{e^{\alpha} (-1 + v^2)^2}\\
- \frac{(1 + K) v}{e^{\alpha} (-1 + v^2)} & \frac{(1 + K) (1 + v^2) \mu}{e^{\alpha} (-1 + v^2)^2}
\end{pmatrix}, \\
    A^{1} &= \frac{\del F}{\del W} = \begin{pmatrix}
- \frac{(1 + K) v}{-1 + v^2} & \frac{(1 + K) (1 + v^2) \mu}{(-1 + \
v^2)^2}\\
\frac{K + v^2}{1 -  v^2} & \frac{2 (1 + K) v \mu}{(-1 + v^2)^2}
\end{pmatrix}.
\end{align*}
The characteristic speeds of \eqref{eqn:Euler_FC_1}-\eqref{eqn:Euler_FC_2} are given by the eigenvalues of the flux Jacobian $\mathbf{F}$
\begin{align}
    \mathbf{F} &= \frac{\del F}{\del V} = \frac{\del F}{\del W}\frac{\del W}{\del V} = A^{1}(A^{0})^{-1}  \\
    &= \begin{pmatrix}
0 & e^{\alpha}\\
\frac{e^{\alpha} (- K + v^2)}{-1 + K v^2} & \frac{2 e^{\alpha} (-1 + K) v}{-1 + K v^2}
\end{pmatrix}. \nonumber 
\end{align}
The characteristic speeds are therefore
\begin{align}
\lambda_{\pm} = \frac{e^{\alpha } \big((-1 + K) v \pm \sqrt{K}(1 - v^2)\big)}{-1 + K v^2}.
\end{align}

\subsection{The Complete System of Equations}
Combining \eqref{eqn:alpha_evo}, \eqref{eqn:A0_FC_evo}-\eqref{eqn:A1_FC_evo}, and \eqref{eqn:Balance_LawSystem} yields the evolution system we numerically solve. These equations can be expressed in matrix form as
\begin{align}
\label{eqn:EinsteinEuler1}
&\del_{t}\mathbf{U} + \del_{x}\mathbf{S} = \mathbf{P}, \\
\label{eqn:EinsteinEuler2}
&\del_{t} \begin{pmatrix} \alpha \\ A \\ U  \end{pmatrix} = \begin{pmatrix} G_{\alpha} \\ A_{0} \\ U_{0}  \end{pmatrix}, \\
\label{eqn:EinsteinEuler3}
&\del_{t}V + \del_{x}F = G
\end{align}
where
\begin{align*}
\mathbf{U} &= e^{-\alpha}(A_{0},A_{1},U_{0}-\frac{1}{2},U_{1})^T, \\ 
\mathbf{S} &= (-A_{1},-A_{0},-U_{1},-U_{0})^T, \\
\mathbf{P} &= e^{-\alpha}\bigl(A_{0} + 4A_{1}U_{1} - 4A_{0}U_{0}, 0 ,\frac{1}{2}e^{-2t+4U}(A_{0}^{2}-A_{1}^{2}) +\frac{1}{2} - U_{0}, 0 \biggr)^T,  \\
G_{\alpha} &= 1 + (K-1)\mu,
\end{align*}
and $V$, $F$, and $G$ are as defined above in Section \ref{sec:conservation_euler}. This system is subject to two constraints that arise from the definitions of $A_{1}$ and $U_{1}$, 
\begin{align}
\label{eqn:A1_constraint}
    C_{A_{1}} \equiv e^{\alpha}\del_{x}A - A_{1} = 0, \\
\label{eqn:U1_constraint}
    C_{U_{1}} \equiv e^{\alpha}\del_{x}U - U_{1} = 0.
\end{align}

\section{FLRW Solutions}
\label{sec:FLRW_Derivation}

The main aim of this article is to study Gowdy-symmetric perturbations of the FLRW solution to the Einstein-Euler system. First, observe that a FLRW metric can be recovered from the Gowdy metric \eqref{eqn:gowdymetricA} by setting $\nu = t + \alpha$, $U=\frac{t}{2}$, $A=0$, and assuming that the remaining metric
function $\alpha$ depends only on $t$. This gives a metric of the form
\begin{align}
g = -e^{t+2\alpha(t)}dt^{2} + e^{t}(dx^{2} + dy^{2} + dz^{2}). 
\end{align}
For the matter variables $\mu$ and $v$ spatial homogeneity and isotropy requires that $v = 0$ and that $\mu$ depend only on $t$. For these choices, the Gowdy-symmetric Einstein-Euler equations \eqref{eqn:Awave}-\eqref{eqn:momentum_constraint}, \eqref{eqn:Balance_LawSystem} simplify to
\begin{align}
\label{eqn:spatially_homog1}
    \del_{t}\mu &= 0, \\
\label{eqn:spatially_homog2}
    \del_{t}\alpha &= 1+(K-1)\mu, \\
\label{eqn:spatially_homog3}
    \mu &= \frac{3}{4}.
\end{align}
Substituting \eqref{eqn:spatially_homog3} into \eqref{eqn:spatially_homog2} and solving for $\alpha$ yields
\begin{align}
    \alpha = \frac{1}{4}\big(1+3K\big)t.
\end{align}
Using the identity \eqref{eqn:mu_defn}, we can then obtain the following expression for the physical density
\begin{align}
    \rho = \frac{3}{4}e^{\frac{-3}{2}(1+K)t}.
\end{align}
The FLRW solution is therefore
\begin{equation}
\begin{aligned}
\label{eqn:FLRW_Soln}
    g &= -e^{\frac{3}{2}(1+K)t}dt^{2} + e^{t}\big(dx^{2} + dy^{2} + dz^{2}\big), \\
    \rho &= \frac{3}{4}e^{\frac{-3}{2}(1+K)t}, \\
    u &= e^{\frac{3}{4}(1+K)t} dt.
\end{aligned}
\end{equation}
\begin{rem}
    It is straightforward to express \eqref{eqn:FLRW_Soln} in terms of standard FLRW coordinates. Defining a new time coordinate $\tau$ by\footnote{In the following we ignore constants which play no role.}
    \begin{align}
        \tau = \int e^{\frac{3}{4}(1+K)t} dt,
    \end{align}
    we obtain the following identity
    \begin{align}
        t = \frac{4\log(\frac{3}{4}(1+K)\tau)}{3(1+K)}.
    \end{align}
    In terms of this time coordinate, the metric \eqref{eqn:FLRW_Soln} takes the form
    \begin{align}
        g = -d\tau^{2} + c\tau^{\frac{4}{3(1+K)}}(dx^{2} + dy^{2} + dz^{2})
    \end{align}
    where $c$ is a positive constant. Ignoring constants, the scale factor is therefore
    \begin{align}
        a(\tau) = \tau^{\frac{2}{3(1+K)}}.
    \end{align}
    This is consistent with the FLRW solution given in \cite{Faraoni:2021}*{\S 2.1}.
\end{rem}

\section{Numerical Results}
\label{sec:NumericalResults}

\subsection{Numerical Setup}
\label{sec:NumericalSetup}
In all our simulations the spatial computational domain is taken to be $[0,2\pi]$ with periodic boundary conditions which is discretised using a uniform grid with $N$ grid points.  Spatial derivatives in the gravitational equations \eqref{eqn:EinsteinEuler1}-\eqref{eqn:EinsteinEuler2} are discretised using second-order central finite difference stencils. As discussed in the introduction, we expect shocks to form in the fluid for suitably slow expansion. To deal with discontinuities numerically, we use the second-order accurate\footnote{Second-order accuracy can only be expected where the solution is smooth. Near discontinuities the solution is expected to be approximately first-order accurate, however it may be worse, see \cite{LeVeque:2002}*{\S 8.7}.} central-upwind Kurganov-Tadmor scheme\footnote{See also \cite{LeVeque:2002} for an introduction to finite volume methods.} \cite{KurganovTadmor:2000} to solve the Euler equations. Our numerical scheme for solving \eqref{eqn:EinsteinEuler3} can be expressed in the following semi-discrete form
\begin{align}
\label{eqn:semi_discrete_BalanceLaw}
    \del_{t}V_{i}(t) = -\frac{1}{\Delta x}\big(\tilde{F}_{i+\frac{1}{2}}(t)-\tilde{F}_{i-\frac{1}{2}}(t)\big) + G_{i}(t),
\end{align}
where $\Delta x$ is the spatial step size and $\tilde{F}_{i+\frac{1}{2}}$ is a consistent numerical flux. At each time step, we proceed as follows
\begin{enumerate}
    \item The primitive variables $\mathbf{u}_{i} = (\mu_{i},v_{i})$ are obtained from the conserved variables using equations \eqref{eqn:Gamma_recovery}-\eqref{eqn:w_recovery}.
    \item The source terms are constructed from the primitive variables.
    \item The primitive variables are reconstructed at each cell-edge and used to construct the numerical fluxes.
\end{enumerate}
We use a slope-limited reconstruction of the primitive variables
\begin{align}
    \mathbf{u}_{i+\frac{1}{2}}^{-} &= \mathbf{u}_{i} + \frac{1}{2}\phi(r_{i})(\mathbf{u}_{i+1}-\mathbf{u}_{i}), \\
    \mathbf{u}_{i+\frac{1}{2}}^{+} &= \mathbf{u}_{i+1} - \frac{1}{2}\phi(r_{i+1})(\mathbf{u}_{i+2}-\mathbf{u}_{i+1}), \\
    r_{i} &= \frac{\mathbf{u}_{i}-\mathbf{u}_{i-1}}{\mathbf{u}_{i+1}-\mathbf{u}_{i}},
\end{align}
where the superscripts $-$ and $+$ denote the reconstruction approaching from the left and right sides of the cell-edge, respectively, and $\phi$ is a slope limiter function, see \cite{LeVeque:2002}*{\S 6.9}. The results presented in this article all use the simple minmod limiter\footnote{Note, this is not the form of the minmod limiter given in \cite{KurganovTadmor:2000} however the method used in this article is equivalent. See the discussion in \cite{LeVeque:1992}*{\S 16.3}}
\begin{align}
    \phi(r) = \max\big(0,\min(r,1)\big),
\end{align}
however we have found comparable results using alternative choices for $\phi$. The numerical fluxes are constructed as follows, 
\begin{align}
    \tilde{F}_{i+\frac{1}{2}} = \frac{1}{2}\Big(F(\mathbf{u}^{+}_{i+\frac{1}{2}}) + F(\mathbf{u}^{-}_{i+\frac{1}{2}}) - a_{i+\frac{1}{2}}\big(\mathbf{u}^{+}_{i+\frac{1}{2}}-\mathbf{u}^{-}_{i+\frac{1}{2}}\big)\Big)
\end{align}
where $a_{i+\frac{1}{2}}(t)$ is the maximum local characteristic speed given by
\begin{align}
    a_{i+\frac{1}{2}}(t) = \max\Big(\varphi\big(\mathbf{F}(\mathbf{u}^{+}_{i+\frac{1}{2}})\big), \varphi\big(\mathbf{F}(\mathbf{u}^{-}_{i+\frac{1}{2}})\big)\Big)
\end{align}
and $\varphi(\mathbf{F})$ denotes the spectral radius of the flux Jacobian. Time integration is performed using a fourth-order Runge-Kutta scheme. To ensure a stable evolution, we set the timestep to be
\begin{align}
    \Delta t = C\frac{\Delta x}{|\lambda_{\max}|}
\end{align}
where $C$ is the CFL constant and $\lambda_{\max}$ is the largest value of the characteristic speeds in the domain.

\subsubsection{Initial Data}
As mentioned previously, the Hamiltonian constraint \eqref{eqn:hamiltonian_constraint} does not need to be imposed on the initial data, however we do need to impose the integral condition \eqref{eqn:integral_condition_ID} and the constraints \eqref{eqn:A1_constraint}-\eqref{eqn:U1_constraint} that arise from the definition of the first order variables $A_{1}$ and $U_{1}$. For the remainder of this section, we use initial data at $t=0$ of the form 
\begin{equation}
\begin{aligned}
\label{eqn:numerical_ID}
    \alpha &= a, \\ 
    U &= c\sin(x), \\
    U_{0} &= \frac{1}{2}+d ,\\
    U_{1} &= e^{\alpha}\del_{x}U, \\
    A &= c\sin(x), \\
    A_{0} &= 0, \\
    A_{1} &= e^{\alpha}\del_{x}A, \\
    \mu &= \frac{3}{4}(1-v^{2}), \\
    v &= b\sin(x),  
\end{aligned}
\end{equation}
where $a$, $b$, $c$, $d$ are constants to be specified with $|b|<1$. Initial data of this form can be considered a perturbation of the FLRW solution \eqref{eqn:FLRW_Soln}, as \eqref{eqn:numerical_ID} reduces to isotropic and homogeneous (i.e. FLRW) initial data when $a=b=c=d=0$. The main aim of this article is to study shock formation resulting from \textit{small} perturbations of the FLRW solution.  In particular, all the plots of this section, with the exceptions of Sections \ref{sec:code_validity} and \ref{sec:shock_time_tests}, are generated using $a=b=c=d=0.05$.

\subsection{Code Tests}
\label{sec:code_tests}
We have verified the accuracy of our code with  convergence tests using resolutions of $N = 400, \; 800, \; 1600, \; 3200$, and $6400$ grid points. The numerical discretisation error $\Delta$ is estimated by taking the $\log_{2}$ of the absolute value of the difference between each simulation and the highest resolution run. \newline \par 

\begin{figure}[htbp]
\centering
\subfigure[Subfigure 1 list of figures text][$t=8.804$.]{
\includegraphics[width=0.45\textwidth]{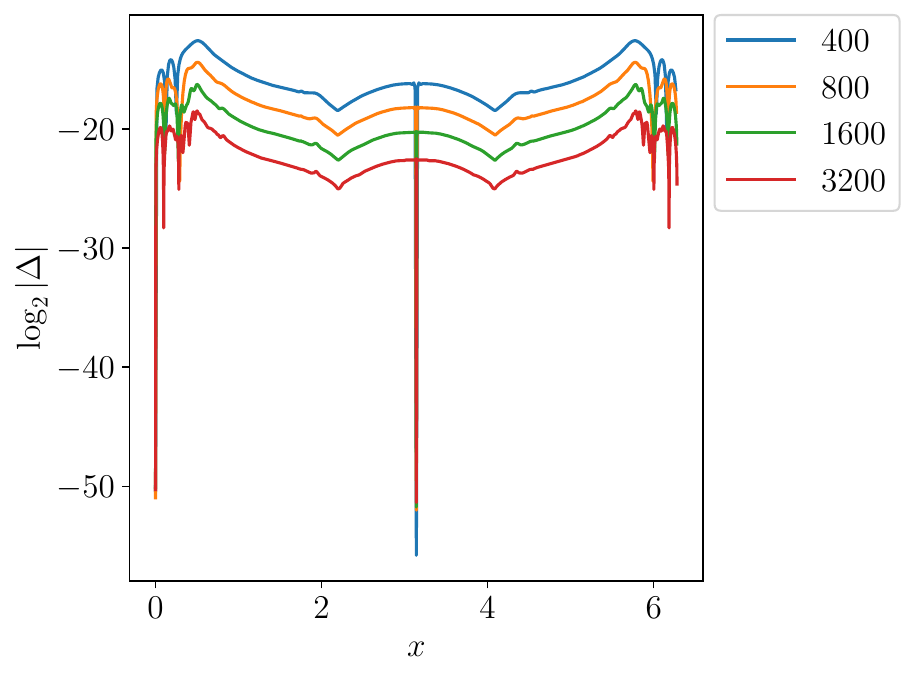}
\label{fig:v_convergence_smooth}}
\subfigure[Subfigure 2 list of figures text][$t=12.14$.]{
\includegraphics[width=0.45\textwidth]{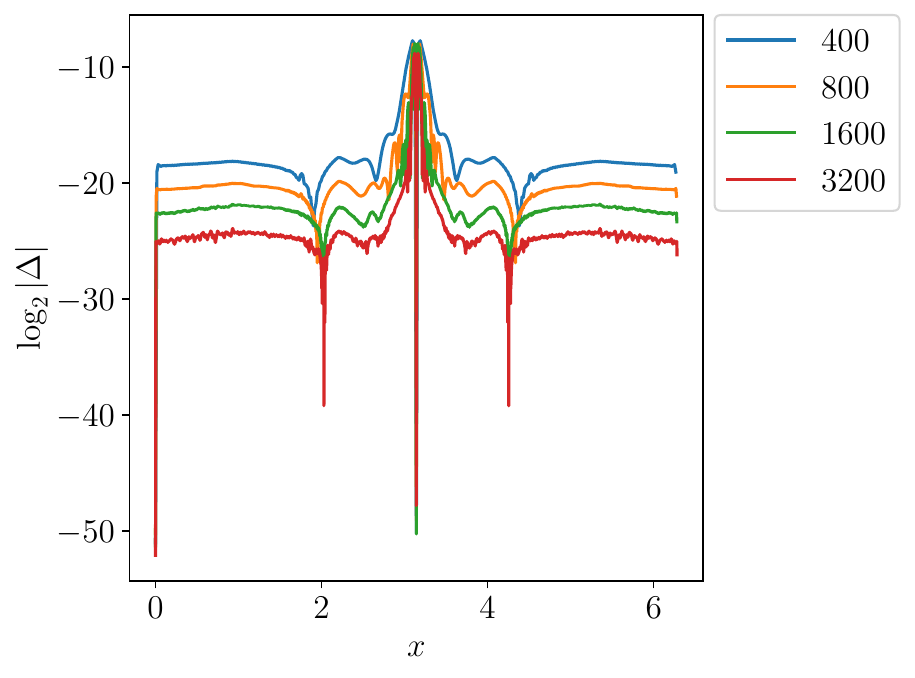}
\label{fig:v_convergence_shocks}}
\caption{Convergence plots of $v$ before and after shocks have formed.}
\end{figure}

It should be noted that the convergence tests we have used are based on Taylor series expansions which implicitly assume the solution is smooth. This assumption does not hold for weak solutions containing discontinuities and thus the expected order of convergence based on the Taylor series will likely not be seen in practice, see for example the discussion in \cite{LeVeque:2002}*{\S 8.7}. Before shocks form we observe the expected second-order convergence in all variables, shown for $v$ in Figure \ref{fig:v_convergence_smooth}. After a shock develops we observe  approximately first order convergence near the shock and slightly worse than second-order convergence elsewhere in the domain. This is shown for $v$ in Figure \ref{fig:v_convergence_shocks}.  Nevertheless, all our simulations are found to converge, even when discontinuities are present. \newline \par 

\begin{figure}
\centering\includegraphics[width=0.5\textwidth]{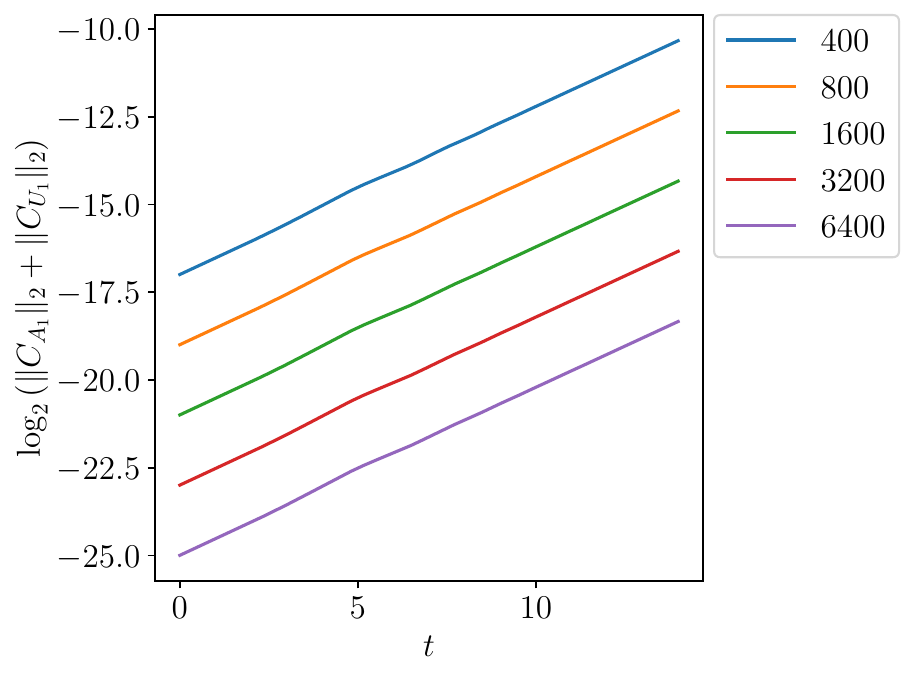}
    \caption{Convergence plot of $\log_{2}\big(\|C_{A_{1}}\|_{2}+\|C_{U_{1}}\|_{2}\big)$, $K=0.1$.}
\label{fig:constraint_sum_convergence}
\end{figure}

There are several other tests we can perform to check the accuracy of our solutions. First, we can monitor the violation of the constraints \eqref{eqn:A1_constraint}-\eqref{eqn:U1_constraint}. Clearly, when 
\begin{align}
    C_{A_{1}}=C_{U_{1}} = 0 
\end{align}
the constraints are identically satisfied. The quantity $\log_{2}\big(\|C\|_{2}\big)$ can
therefore be understood as the violation error of the constraint as a function of time. We observe approximately second order convergence throughout the simulation.  This is shown for the combined constraint quantity $\log_{2}\big(\|C_{A_{1}}\|_{2}+\|C_{U_{1}}\|_{2}\big)$ in Figure \ref{fig:constraint_sum_convergence}.

\begin{figure}
    \centering
    \includegraphics[width=0.5\textwidth]{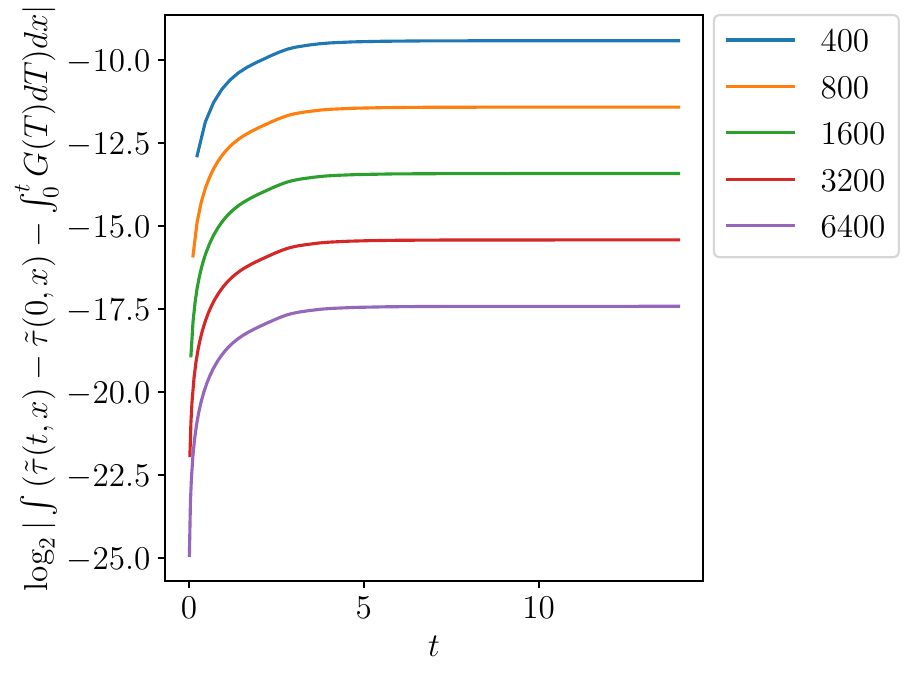}
    \caption{Convergence of $\int_{\mathcal{D}} \big(\tilde{\tau}(t,x)-\tilde{\tau}(0,x) - \int_{0}^{t} G(T) dT \big) dx$, $K=0.1$.}
    \label{fig:Balance_Law_Conservation_Converge}
\end{figure}

Let us now consider the integral form of a general conservation law on a domain $\mathcal{D}$ without boundaries,
\begin{align}
    \frac{d}{dt}\int_{\mathcal{D}} u dx + \int_{\mathcal{D}} \del_{x}f = 0.
\end{align}
By applying the divergence theorem, the integral of the flux disappears and we are left with
\begin{align}
\label{eqn:conserved_variable}
    \frac{d}{dt}\int_{\mathcal{D}} u dx = 0.
\end{align}
A finite volume scheme should preserve \eqref{eqn:conserved_variable} up to machine precision. For a balance law (i.e. with source terms)
\begin{align}
\label{eqn:balance_law_example}
    \del_{t}V + \del_{x}F = G,
\end{align}
the equivalent condition is
\begin{align}
\label{eqn:conserved_balance_law}
    \int_{\mathcal{D}} \Big(V(t,x)-V(0,x) - \int_{0}^{t}G(T,x) dT\Big) dx = 0.
\end{align}
In general, we should only expect our scheme to preserve \eqref{eqn:conserved_balance_law} up to truncation error\footnote{Note, however, that one can employ so-called well-balanced schemes which preserve certain steady states to much higher accuracy, see for example \cite{LeFloch:2021_wellbalanced}.}. We have checked this numerically, using the trapezoid rule to evaluate the integral of the source term. Our scheme achieves the expected convergence rate, shown for $\tilde{\tau}:=e^{-\alpha}\tau$ in Figure \ref{fig:Balance_Law_Conservation_Converge}. 

\subsubsection{Code Validation}
\label{sec:code_validity}
\begin{figure}
\centering\includegraphics[width=0.5\textwidth]{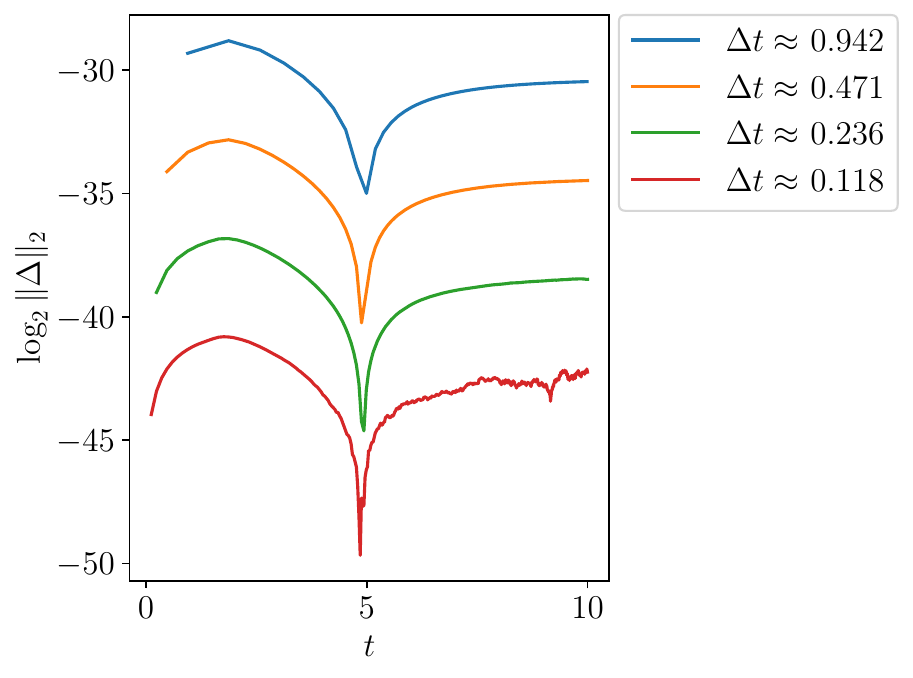}
    \caption{Convergence plot of $L_{2}$-norm of $\alpha-\alpha_{\text{exact}}$ for several different values of the timestep $\Delta t$. All evolutions used $N=200$ and $K=0.25$.}
\label{fig:FLRW_Convergence}
\end{figure}

As a further code check, we can test that our code correctly replicates the exact FLRW solution \eqref{eqn:FLRW_Soln}. We use the following initial data, which is obtained by setting $a=b=c=d=0$,
\begin{equation}
\begin{aligned}
\label{eqn:homog_ID}
\mu &= \frac{3}{4}, \\
\alpha &= \nu = U = U_{1} = U_{0} = A = A_{1} = A_{0} = v = 0.
\end{aligned}
\end{equation}
As the solution is spatially homogeneous, the order of convergence is determined solely by our time-stepping scheme. Our scheme displays the expected fourth-order convergence, shown for $\alpha$ in Figure \ref{fig:FLRW_Convergence}.

\subsection{Testing for Shock Formation}
Our numerical scheme, unlike that of \cite{Fajman_et_al:2024}, uses shock capturing methods which allows us to evolve beyond the point of shock formation. Typically, it is clear from visual inspection that a shock has formed in the fluid. As a further check, and to maintain consistency with \cite{Fajman_et_al:2024}, we can also monitor the norm of our fluid variables. \newline \par

\begin{figure}
\centering\includegraphics[width=0.5\textwidth]{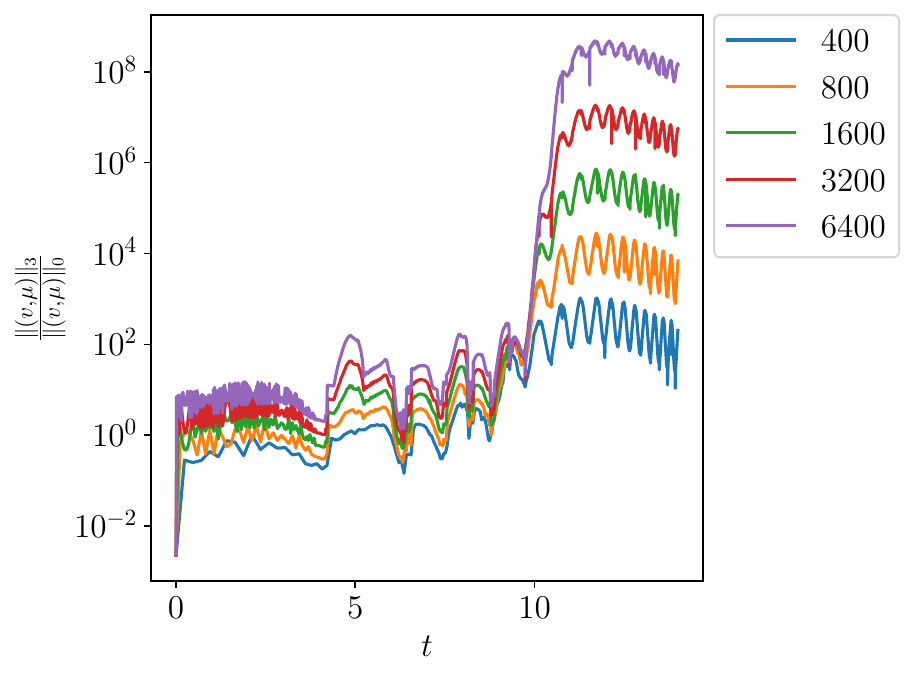}
    \caption{Values of the norm ratio \eqref{eqn:EM_NormRatio} over time for various resolutions, $K=0.1$. Shocks form just after $t=10$.}
\label{fig:Norm_RatioPlots}
\end{figure}

In \cite{Fajman_et_al:2024}, the formation of shocks was determined by monitoring the ratio of norms
\begin{align}
\label{eqn:Fajman_NormRatio}
    \frac{\|(v,L)\|_{3}}{\|(v,L)\|_{0}}
\end{align}
where
\begin{align}
    \|(v,L)\|_{k} = \|\del^{k}_{x}v\|_{L^{2}}^{2} + \|\del^{k}_{x}L\|_{L^{2}}^{2},
\end{align}
and $L$ is the re-scaled density variable used in \cite{Fajman_et_al:2024}. As the solution becomes discontinuous, the ratio \eqref{eqn:Fajman_NormRatio} should blow up. In \cite{Fajman_et_al:2024}, a solution was classified as shock-forming if this ratio was larger than $10^{6}$. In our scheme, we monitor the quantity
\begin{align}
\label{eqn:EM_NormRatio}
    \frac{\|(v,\mu)\|_{3}}{\|(v,\mu)\|_{0}},
\end{align}
using a fourth-order finite difference stencil to approximate the third derivative. There is an obvious downside to using a ratio of norms, such as \eqref{eqn:EM_NormRatio}, to determine if a shock has formed as the size of this quantity is \textit{resolution-dependent}. As the number of gridpoints is increased, steep gradients in the numerical solution will produce a larger value of the norm ratio. To mitigate this effect, we monitor its size for a sequence of solutions with increasing resolution. This is shown in Figure \ref{fig:Norm_RatioPlots}. We find that a combination of monitoring \eqref{eqn:EM_NormRatio} and evolving suitably past the predicted time of shock formation allows to us consistently identify the presence of shocks. 

\subsection{Numerical Behaviour}
\subsubsection{Shock Formation for $K>0$}
\begin{figure}[htbp]
\centering
\subfigure[Subfigure 1 list of figures text][$t = 0$]{
\includegraphics[width=0.3\textwidth]{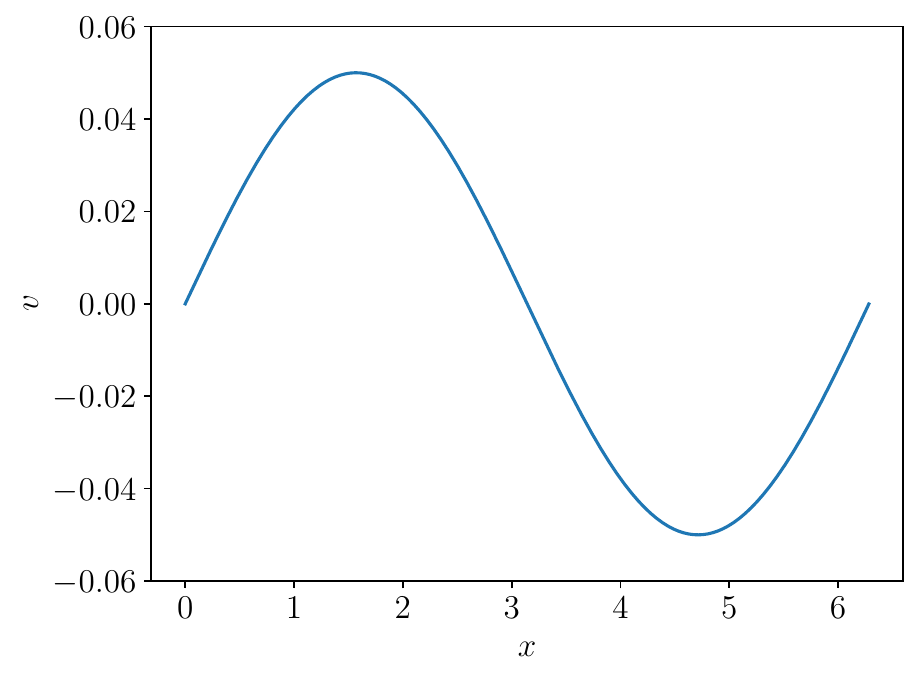}
\label{fig:subfigv_t0}}
\subfigure[Subfigure 2 list of figures text][$t = 7.39$]{
\includegraphics[width=0.3\textwidth]{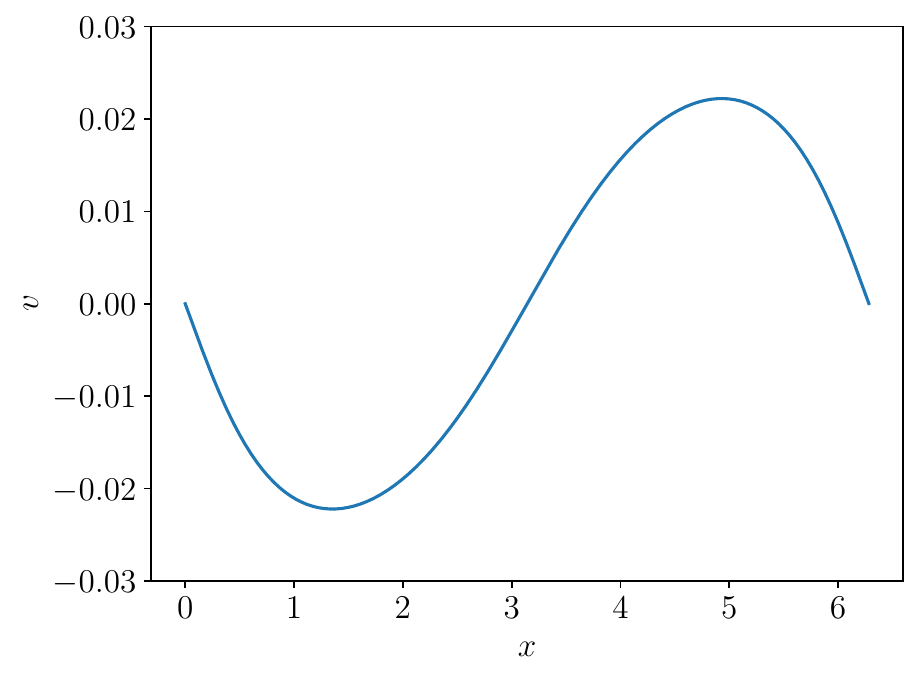}
\label{fig:subfigv_t7_45}}
\subfigure[Subfigure 2 list of figures text][$t = 11.58$]{
\includegraphics[width=0.3\textwidth]{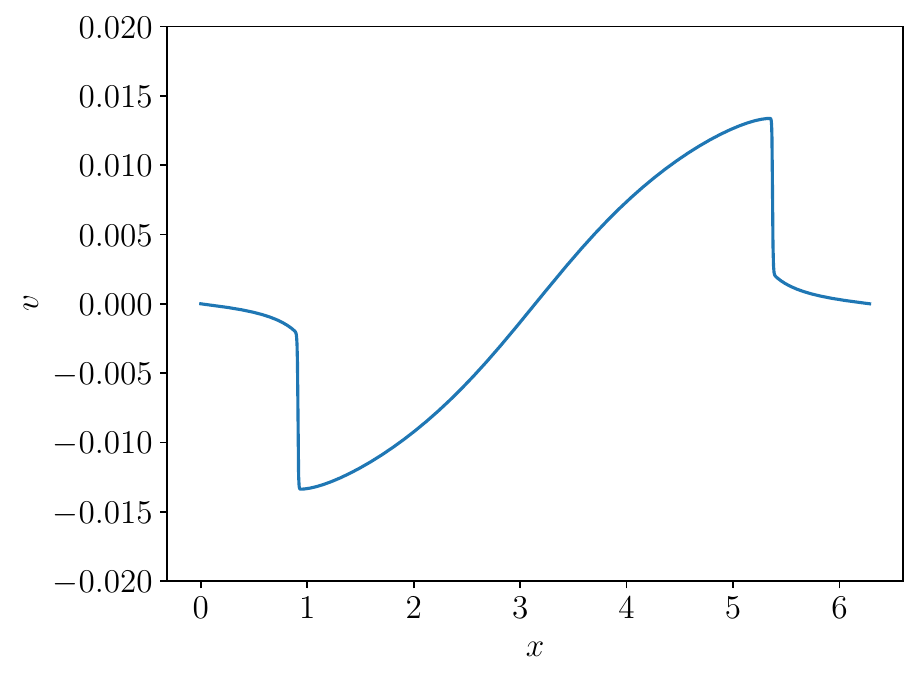}
\label{fig:subfigv_t11_58}}
\caption{Fluid velocity $v$ at various times. $N=6400$, $K=0.1$. Two shocks are clearly present in Figure \ref{fig:subfigv_t11_58}.}
\label{fig:Shock_Formation_K01}
\end{figure}

\begin{figure}
\centering\includegraphics[width=0.5\textwidth]{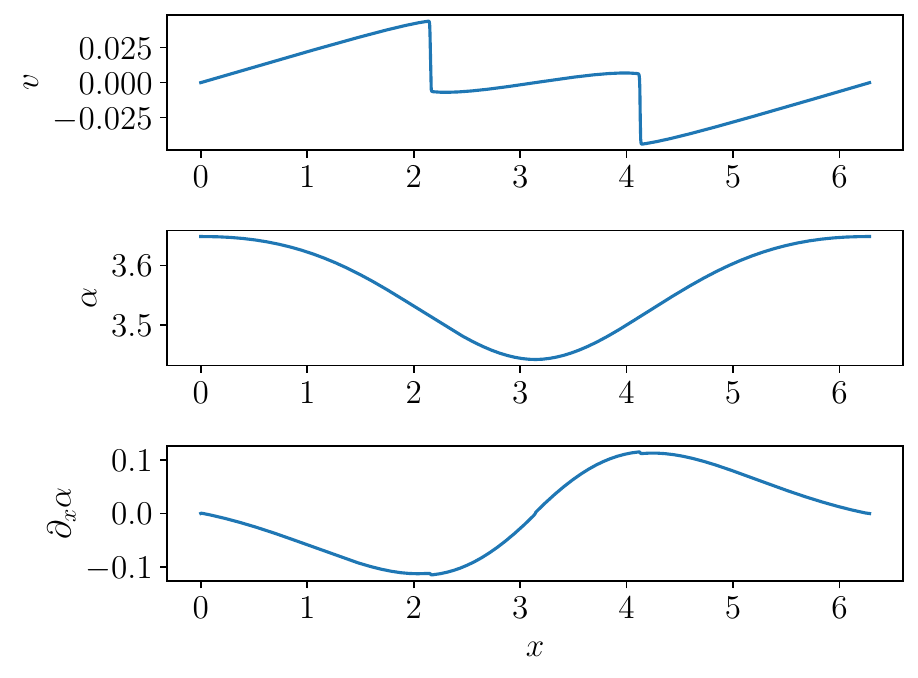}
    \caption{Fluid velocity $v$ (top), $\alpha$ (middle), and $\del_{x}\alpha$ (bottom) at $t=7.41$. The derivative of $\alpha$ was estimated using second order finite differences. $N=6400$, $K=0.3$.}
\label{fig:alpha_deriv_comparison}
\end{figure}

\begin{figure}
\centering\includegraphics[width=0.5\textwidth]{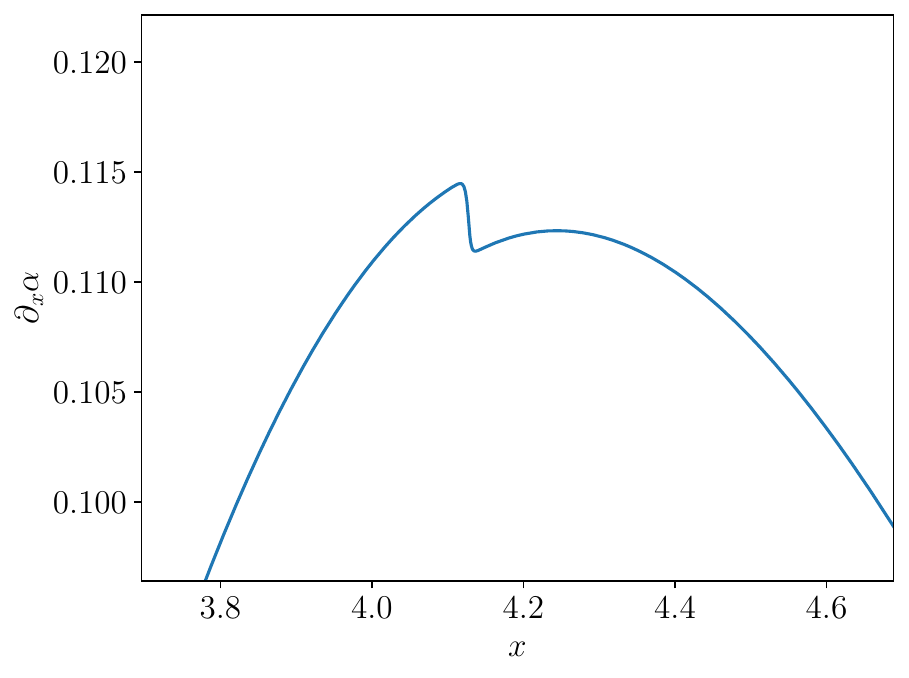}
    \caption{Close up of kink in $\del_{x}\alpha$ at $t=7.41$. $N=6400$, $K=0.3$.}
\label{fig:alpha_deriv_kink_closeup}
\end{figure}

We now discuss the behaviour of numerical solutions generated from initial data of the form \eqref{eqn:numerical_ID} for $K \in (0,1]$. As discussed in the introduction, we observe that small perturbations of the FLRW solution always develop shocks in the fluid variables in finite time. This is shown for the fluid velocity $v$ in Figure \ref{fig:Shock_Formation_K01}. We do not observe discontinuities in any of the gravitational variables even after shocks form in the fluid. It should be noted that the evolution equation for $\alpha$ \eqref{eqn:alpha_evo}, unlike the other gravitational variables, includes the fluid density $\mu$ which can (and does) contain jump discontinuities. Indeed, we find that the derivative $\del_{x}\alpha$ contains `kinks' at the same spatial position as the shocks in the fluid variables. This is shown in Figures \ref{fig:alpha_deriv_comparison} and \ref{fig:alpha_deriv_kink_closeup}. \newline \par

In general, the fluid variables behave similarly to unstable fluids in the fixed background case studied in \cite{Fajman_et_al:2024}\footnote{We have also performed our own tests of the fixed background problem using the shock capturing method described in Section \ref{sec:NumericalSetup} and observed the same behaviour described in \cite{Fajman_et_al:2024}.}. In particular, the fluid variables are found to oscillate from side to side at an increasing frequency with shocks eventually forming in the fluid velocity and density. Similar oscillations are seen in the metric variables, however as mentioned before shocks do not form. This behaviour is consistent with previous analytical work in the linear and decelerated regimes \cites{Fajman_et_al:2025,FOOW:2023,FOW:2021}.  However, this is completely different to the asymptotic behaviour for perturbed FLRW spacetimes with accelerated expansion which become ODE-dominated at late times, see for example \cites{BMO:2023,Oliynyk:CMP_2016, RodnianskiSpeck:2013, Speck:2013, HadzicSpeck:2015,LubbeKroon:2013, Fournodavlos_et_al:2024}. \newline \par

It should be noted that at the endpoint $K=1$, the frequency of the oscillations becomes \textit{significantly} higher. This seems to be significantly more challenging for our scheme to evolve and errors while recovering the primitive variables eventually lead to our code crashing. However, based on the norm ratio, it appears that shocks are still forming in the solution before the code crashes.

\begin{rem}
    In \cite{Rendall:2004}, Rendall conjectured that super-radiative ($K>1/3$) FLRW spacetimes with \textit{accelerated} expansion would be unstable towards the future in the sense that the fractional density gradient $\frac{\del_{x}\rho}{\rho}$ of the fluid will blow-up at future timelike infinity. In the articles \cites{BMO:2023,BMO:2024,Oliynyk:2024} it was established that this `Rendall instability' is driven by the difference in asymptotic behaviour of tilted and orthogonal homogeneous fluids. In particular, this is because cosmological models with accelerated expansion are well approximated (pointwise) by homogeneous solutions at late times. It is interesting, therefore, to note that we \textit{never} observe the Rendall instability for perturbations of decelerated FLRW models with $K>\frac{1}{3}$. Indeed, we observe  little qualitative difference between simulations with sub-radiative and super-radiative equations of state. We believe this is because perturbed models in the decelerated regime with $K>0$ are \textit{never} ODE dominated at late times and, thus, the homogeneous behaviour of fluids does not play a significant role in the late time asymptotics of these solutions. 
\end{rem}

\subsubsection{Scaling of Shock Formation Time for $K>0$}
\label{sec:shock_time_tests}
The shock formation time for barotropic fluids ($K>0$) is observed to decrease as both $K$ and the size of our perturbations are increased. In order to investigate this scaling, we have tested two cases:
\begin{enumerate}
    \item The size of the initial fluid perturbation, $b$, is a free parameter, $a=c=d=0.05$ and $K>0$ is fixed.
    \item $0 < K \leq 1$ is a free parameter and $a=b=c=d=0.05$.
\end{enumerate}
For both of these cases, the system \eqref{eqn:EinsteinEuler1}-\eqref{eqn:EinsteinEuler3} was evolved using $6400$ grid points. The shock formation time was estimated to be the first time that the norm ratio \eqref{eqn:EM_NormRatio} exceeded $10^{6}$. \newline \par

\begin{figure}
\centering\includegraphics[width=0.5\textwidth]{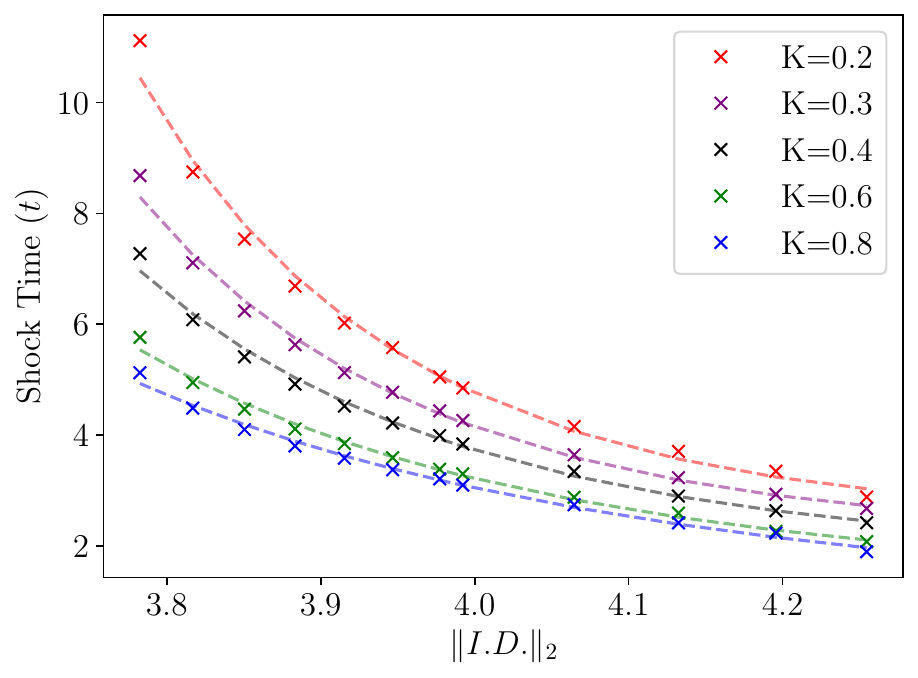}
    \caption{Plot of the estimated shock formation time against the size of initial data, $\|I.D.\|_{2}$ for various values of $K$. The dashed lines are curves of best fit following the formula \eqref{eqn:approximate_data_scaling}. Values of $b$ were taken between $0.02$ and $0.35$. $N=6400$, $a=c=d=0.05$.}
\label{fig:ShockTime_IDScaling}
\end{figure}

\begin{figure}
\centering\includegraphics[width=0.5\textwidth]{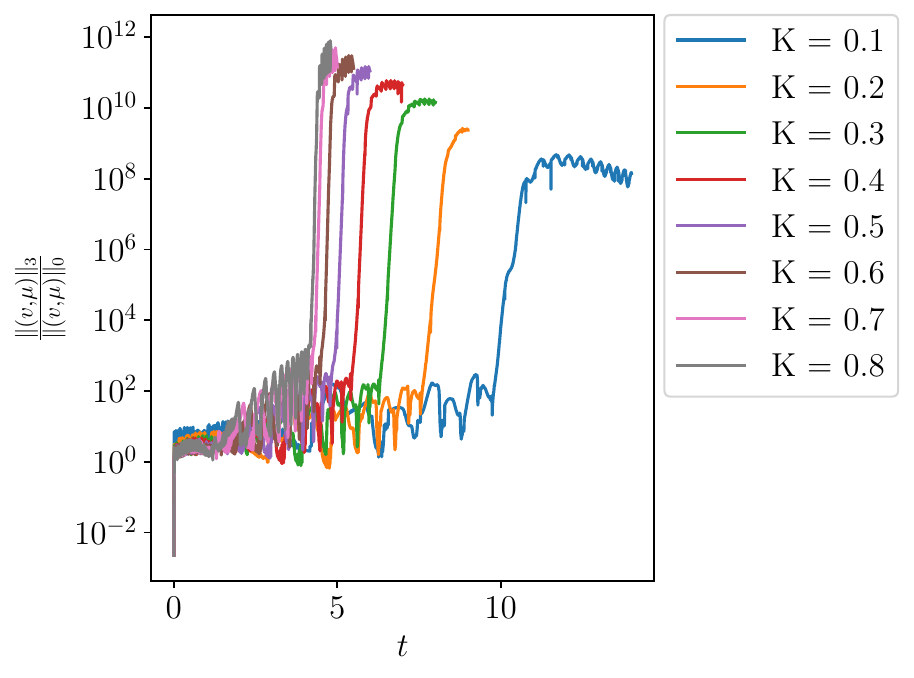}
    \caption{Values of the norm ratio \eqref{eqn:EM_NormRatio} over time for various values of $K$. $N=6400$ and $a=b=c=d=0.05$.}
\label{fig:Norm_Ratio_KComparison}
\end{figure}

\begin{figure}
\centering\includegraphics[width=0.5\textwidth]{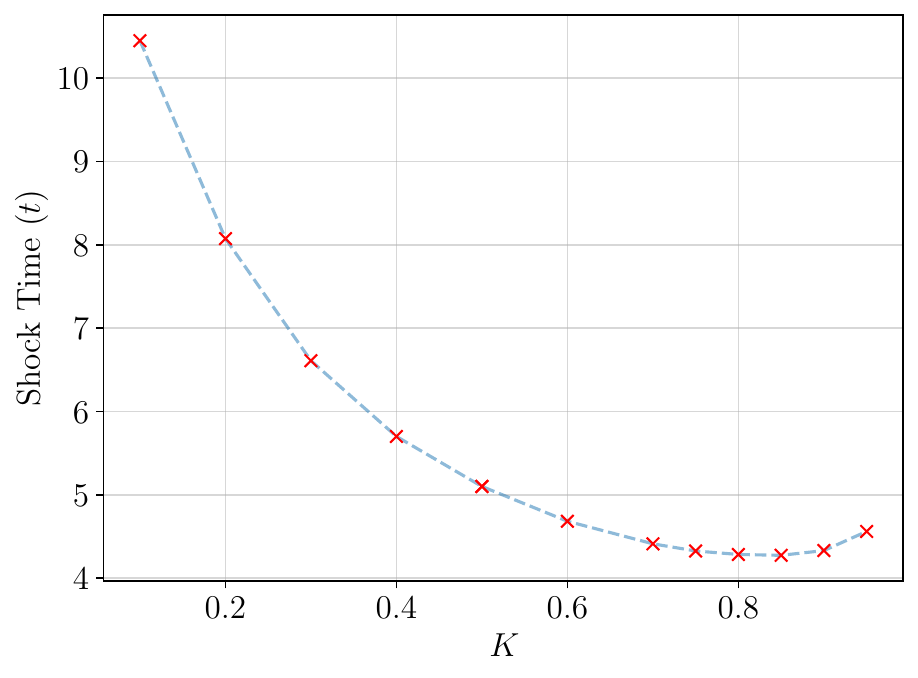}
    \caption{Estimated shock formation time for different values of $K$. $N=6400$ and $a=b=c=d=0.05$.}
\label{fig:ShockTime_KScaling}
\end{figure}

As the initial fluid perturbation $b$ is increased, we find the shock time decreases exponentially. In particular we find, for fixed $K$, the shock formation time $t_{*}$ scales approximately as 
\begin{align}
\label{eqn:approximate_data_scaling}
    t_{*} \sim A\exp\Big(\frac{B}{\|I.D.\|_{2}^{2}} + \frac{C}{\|I.D.\|_{2}}\Big)
\end{align}
where $A$, $B$, and $C$ are constants and $\|I.D.\|_{2}$ is the $L_{2}$ norm of the initial data
\begin{align}
    \|I.D.\|_{2} &:= \|v(0,x)\|_{2} + \|\mu(0,x)\|_{2} + \|A(0,x)\|_{2} + \|A_{0}(0,x)\|_{2} + \|A_{1}(0,x)\|_{2} \nonumber \\
    &+ \|U(0,x)\|_{2} + \|U_{0}(0,x)\|_{2} + \|U_{1}(0,x)\|_{2} + \|\alpha(0,x)\|_{2}.
\end{align}
This is shown in Figure \ref{fig:ShockTime_IDScaling} for different values of $K$. Similarly, for case (2),  the shock time is observed to decrease exponentially as $K$ increases before plateauing around $K=0.8$. Unexpectedly, however, the shock time begins to increase again for $K\geq 0.9$. This is shown in Figures \ref{fig:Norm_Ratio_KComparison} and \ref{fig:ShockTime_KScaling}. 

\subsubsection{Shock Formation for $K=0$}
As discussed in the introduction, the behaviour of fluids can be significantly different when $K=0$. In particular, dust is stable when evolved on fixed power-law FLRW-type backgrounds for a much greater range of $\sigma$ than barotropic fluids ($K>0)$. Recent work \cite{Ju_et_al:2025} has also established that, in addition to shocks forming in the fluid velocity, the density itself may blow up in finite time for dust fluids on fixed FLRW backgrounds. \newline \par

In our simulations, we find the oscillatory behaviour observed in the fluid variables for $K>0$ is no longer present. Instead, the velocity forms a single stationary shock, which contrasts the behaviour observed in simulations of the fixed FLRW background case where shocks do \textit{not} form. The density variable $\mu$ develops sharp features however it remains bounded and does not appear to form shocks. Oscillatory behaviour is still present in the gravitational variables, although at a significantly reduced frequency. \newline \par

We wish to emphasise that our numerical scheme becomes significantly more unstable when $K=0$. We find sharp, seemingly discontinuous, features form in certain metric variables before spurious high frequency oscillations\footnote{Not to be confused with the oscillations seen in the $K>0$ setting.} set in and the code crashes. Significantly reducing the CFL number and evolving the variables $A_{0}$, $A_{1}$, $U_{0}$, and $U_{1}$ with the Kurganov-Tadmor method was found to  stabilise the scheme, however the spiky features in metric variables remained present. Given these numerical issues, further simulations of the dust case with an alternative numerical scheme would be needed to confidently assert the presence of shocks.   \newline \par

\begin{figure}[htbp]
\centering
\subfigure[Subfigure 1 list of figures text][$v$]{
\includegraphics[width=0.45\textwidth]{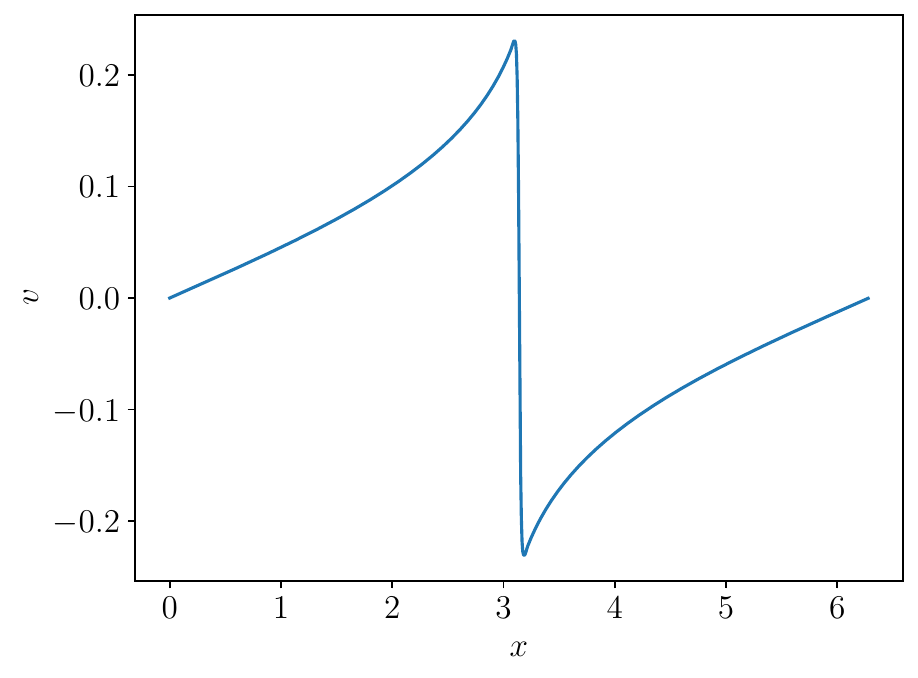}
\label{fig:subfigv_K0}}
\subfigure[Subfigure 2 list of figures text][$U_{1}$]{
\includegraphics[width=0.45\textwidth]{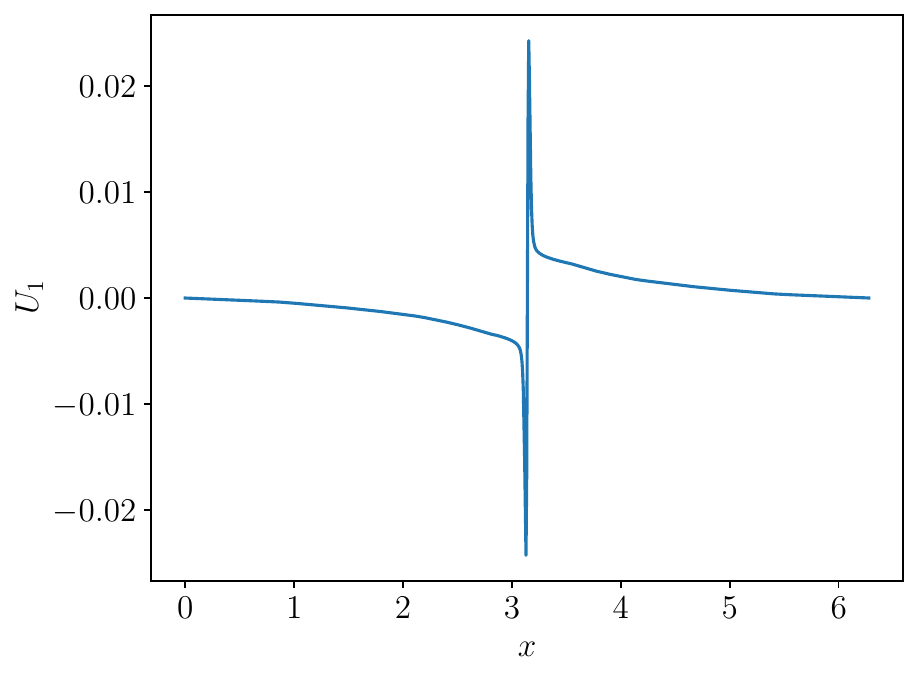}
\label{fig:subfigA_K0}}
\caption{$v$ and $U_{1}$ at $t=6.59$. $N=1000$, $K=0$. The Kurganov-Tadmor scheme was used to evolve $A_{0}$, $A_{1}$, $U_{0}$, and $U_{1}$ for this simulation.}
\label{fig:plots_K0}
\end{figure}

It is unclear whether the sharp features observed in the metric functions are numerical errors, genuine discontinuities, or merely something similar to a `spike'  (see for example \cites{Berger:1993,Garfinkle:2004,ColeyLim:2015, berger1998c,RendallWeaver:2001, GarfinklePretorius:2020, BMO:2024}) which our code is not sufficiently resolving. This behaviour is shown for $U_{1}$, along with a shock in the fluid velocity, in Figure \ref{fig:plots_K0}. 

\section{Discussion}
We have numerically simulated Gowdy-symmetric nonlinear perturbations of FLRW solutions to the Einstein-Euler system for the full sound speed range $0\leq K \leq 1$ using shock-capturing methods. As shocks are thought to play an important role during structure formation in the early universe \cites{LiTurok:2016,Miniati_et_al:2000,Ryu_et_al:2003}, it is of physical interest to determine the conditions under which they form in near-FLRW spacetimes. In particular, we have presented strong evidence that arbitrarily small perturbations of the \textit{decelerated} FLRW solution form shocks in finite time.  For $K \in (0,1]$, we observe that perturbations of the FLRW solution become highly oscillatory and rapidly develop shocks in the fluid variables. Additionally, we find that the shock formation time decreases exponentially as both the size of the perturbation and $K$ are increased. The behaviour of these solutions is broadly consistent with the results for unstable fluids on fixed decelerating FLRW background spacetimes in \cite{Fajman_et_al:2024}. On the other hand, when $K=0$, the fluid variables no longer oscillate and oscillations in the gravitational variables are significantly damped. The fluid velocity forms a single stationary shock which contrasts the behaviour of pressureless fluids on fixed FLRW backgrounds. In this regime, discontinuities also appear to form in the gravitational variables. Numerical issues for these simulations, however, mean further investigation of perturbed dust spacetimes is required. It should be noted that all of these results starkly contrast the behaviour of spacetimes with \textit{accelerated} expansion where it is known that shock formation in fluids is suppressed for small perturbations \cites{RendallBrauerReula:1994,Speck:2013,RodnianskiSpeck:2013,HadzicSpeck:2015,LubbeKroon:2013,Oliynyk:CMP_2016,Fournodavlos_et_al:2024}. 
 There are many interesting avenues for future research. An obvious step would be to study shock formation without the Gowdy symmetry assumption. Additionally, Taylor \cite{Taylor:2023} has recently proven stability of the FLRW solution to the Einstein-massless Vlasov system on non-compact spatial slices. It would be interesting to determine whether such a stability result also holds for perfect fluids. A natural first step in this direction would be to numerically study perturbations of the FLRW solution in spherical symmetry. We plan on investigating this in a future work. 

\subsection*{Acknowledgements}
I would like to thank Florian Beyer for helpful discussions and feedback on an early draft of this article.

\bibliography{refs.bib}

\end{document}